\documentclass{emulateapj}

\def\kms{\ifmmode{\rm km\thinspace s^{-1}}\else km\thinspace s$^{-1}$\fi}
\def\phiher{$\phi$~Her}

\journalinfo{To appear in The Astronomical Journal, June 2007 issue}
\submitted{}

\shortauthors{Torres}
\shorttitle{\phiher}

\begin{document}

\title{Astrometric-spectroscopic determination of the absolute masses
of the H\lowercase{g}M\lowercase{n} binary star $\phi$~Herculis}

\author{Guillermo Torres}

\affil{Harvard-Smithsonian Center for Astrophysics, 60 Garden St.,
Cambridge, MA 02138}

\email{gtorres@cfa.harvard.edu}

\begin{abstract} 

The Mercury-Manganese star \phiher\ is a well known spectroscopic
binary that has been the subject of a recent study by
\cite{Zavala:06}, in which they resolved the companion using
long-baseline interferometry. The total mass of the binary is now
fairly well established, but the combination of the spectroscopy with
the astrometry has not resulted in individual masses consistent with
the spectral types of the components. The motion of the center of
light of \phiher\ was clearly detected by the {\it Hipparcos\/}
satellite. Here we make use of the {\it Hipparcos\/} intermediate data
(`abscissa residuals') and show that by combining them in an optimal
fashion with the interferometry the individual masses can be obtained
reliably using only astrometry. We re-examine and then incorporate
existing radial-velocity measurements into the orbital solution,
obtaining improved masses of $3.05 \pm 0.24$~M$_{\sun}$ and $1.614 \pm
0.066$~M$_{\sun}$ that are consistent with the theoretical
mass-luminosity relation from recent stellar evolution models. These
mass determinations provide important information for the
understanding of the nature of this peculiar class of stars.
	
\end{abstract}

\keywords{binaries: general --- binaries: spectroscopic --- methods:
data analysis --- stars: chemically peculiar --- stars: fundamental
parameters --- stars: individual (\phiher)}

\section{Introduction}
\label{sec:introduction}

In a recent paper \cite{Zavala:06} reported interferometric
observations of the Mercury-Manganese star \phiher\ (HD~145389,
HR~6023, HIP~79101, $\alpha_{\rm J2000} = 16^{\rm h} 08^{\rm m}
46\fs18$, $\delta_{\rm J2000} = +44\arcdeg 56\arcmin 05\farcs7$,
spectral type B9:p(HgMn), $V = 4.22$), which is a binary star. This
object belongs to a class of peculiar non-magnetic B-type stars that
show abundance anomalies of several elements, some of which (such as
Hg and Mn) can be enhanced by orders of magnitude
\citep{Preston:74}. Depletions of other elements are seen as
well. These anomalies are thought to be produced by radiatively-driven
diffusion and gravitational settling \citep{Michaud:70}. The
observations of \cite{Zavala:06} spatially resolve the companion of
\phiher\ for the first time. The object had previously been known as a
single-lined spectroscopic binary with a period of about 560 days
\citep{Babcock:71, Aikman:76} and an eccentric orbit. Their
interferometric measurements with the Navy Prototype Optical
Interferometer (NPOI) allowed \cite{Zavala:06} to measure the
brightness of the companion and use it to estimate its spectral
type. This, in turn, facilitated their detection of spectral lines of
the secondary for the first time in spectra taken at the Dominion
Astrophysical Observatory. The secondary was also detected
spectroscopically by \cite{Dworetsky:06}.

The system is therefore technically now double-lined, although as
pointed out by \cite{Zavala:06} the handful of secondary radial
velocities they were able to measure with great difficulty do not
provide a firm constraint on the velocity semiamplitude of that
star. Furthermore, they found that the combination of their
high-precision interferometric orbit with the elements of the
single-lined spectroscopic orbit reported by \cite{Aikman:76} led to
absolute masses for the components that are inconsistent with the
spectral types. Therefore, although the total mass of the binary is
now fairly well known, the individual masses cannot yet be determined
dynamically.  Such basic properties of the stars are of considerable
interest given the chemical peculiarities of this class of objects,
which have been the subject of extensive studies of many different
kinds \citep[see, e.g.,][and references therein]{Adelman:01,
Adelman:03, Adelman:04, Dolk:03, Dworetsky:00, Leushin:95}.

We were puzzled by this apparent inconsistency between the seemingly
precise radial velocities and the astrometry, and we wondered whether
other velocity measurements available in the literature might clarify
the situation.  In addition, \phiher\ was observed in the course of
the {\it Hipparcos\/} mission \citep{ESA:97}. Those measurements
clearly revealed the motion of the center of light due to the binary
orbit, suggesting they could be used to good advantage in the
determination of the individual masses.  The motivation for this paper
is therefore threefold: \emph{i)} To show how, in the absence of
spectroscopy, the astrometric measurements from {\it Hipparcos\/} can
indeed be combined with the NPOI observations of \cite{Zavala:06} to
yield reliable masses for both stars for the first time, purely
astrometrically. This serves as an interesting example of the value of
the {\it Hipparcos\/} intermediate data for solving or improving
binary orbits and inferring other stellar properties; \emph{ii)} To
readdress the issue of the \cite{Aikman:76} velocities in light of the
new solution. We show that those data are not really inconsistent with
the astrometry, but do lack critical phase coverage and must be
supplemented by other information in order to be useful; \emph{iii)}
To incorporate additional velocity measurements not previously used,
in order to further strengthen the orbital solution. The final masses
not only have much improved precision, but are consistent with the
mass-luminosity relation as given by current stellar evolution models.

\section{Interferometric observations}
\label{sec:interferometry}

The original NPOI measurements obtained by \cite{Zavala:06} consist of
interferometric squared visibilities ($V^2$) and closure phases
collected on 25 separate nights from 1997 April to 2005 July.  The
visibilities of each night were combined to infer the angular
separation ($\rho$) and position angle ($\theta$) of the binary, and
these were subsequently used to derive the elements of the astrometric
orbit. The measurements published are $\rho$ and $\theta$, along with
the corresponding error ellipses. The residuals from the orbit are
only a few tenths of a milli-arc second (mas) in angular separation,
and a few tenths of a degree in position angle, indicating the very
high precision of those measurements. The orbit has an angular
semimajor axis of 32.1~mas, which, when combined with the {\it
Hipparcos\/} parallax ($\pi_{\rm Hip} = 14.27 \pm 0.52$ mas) allowed
\cite{Zavala:06} to infer the total mass of the binary as $M_{\rm tot}
= 4.7 \pm 0.6$~M$_{\sun}$.  Additionally, the interferometric
measurements yielded the average magnitude difference between the
primary and secondary of \phiher\ at two wavelengths: $\Delta m$
(5500~\AA) $ = 2.57 \pm 0.05$ mag and $\Delta m$ (7000~\AA) $ = 2.39
\pm 0.05$ mag. As discussed below this information is key to the mass
determinations.

\section{{\it Hipparcos\/} observations}
\label{sec:hipparcos}

The star \phiher\ was observed by the {\it Hipparcos\/} satellite
during its 37-month astrometric mission under the designation
HIP~79101, and was measured a total of 76 times from 1989 December to
1993 March, or a slightly over two orbital cycles of the binary. Each
measurement consisted of a one-dimensional position (`abscissa') along
a great circle representing the scanning direction of the satellite,
tied to an absolute frame of reference known as the International
Celestial Reference System (ICRS).  In addition to providing the
trigonometric parallax as well as the position and proper motion of
the center of light based on the synthesis of all these measurements
(the so-called standard 5-parameter fit), the astrometric solution
reported in the Catalogue \citep{ESA:97} revealed a perturbation that
was modeled as orbital motion, since the binary nature of the object
was already known. Several of the orbital elements including the
period ($P$), the eccentricity ($e$), and the longitude of periastron
of the primary ($\omega_{\rm A}$) were held fixed at the values
determined spectroscopically by \cite{Aikman:76} to facilitate the
solution, and the other elements were solved for. These included the
semimajor axis of the center of light ($a_{\rm phot} = 9.09 \pm 0.65$
mas), the inclination angle ($i = 36\arcdeg \pm 14\arcdeg$), the
position angle of the ascending node ($\Omega = 188\arcdeg \pm
12\arcdeg$, J2000), and the time of periastron passage ($T$). More
recently \cite{Jancart:05} reported a reanalysis of the {\it
Hipparcos\/} intermediate data, still adopting $P$, $e$, $\omega_{\rm
A}$, as well as $T$ from the spectroscopic work by \cite{Aikman:76},
but with the added assumption that the secondary contributes no light
to the system. This provides a connection between the measured
semimajor axis $a_{\rm phot}$ and the radial velocity semiamplitude
$K_{\rm A}$, which was adopted also from \cite{Aikman:76}. Their
results for the inclination ($i = 10\fdg3 \pm 0\fdg7$) and especially
the position angle of the node ($\Omega = 148\fdg3 \pm 3\fdg0$) are
quite different from the {\it Hipparcos\/} solution, but we now know
from the detection of the secondary by \cite{Zavala:06} that this star
\emph{does} contribute some light, so the assumption of
\cite{Jancart:05} is not a good one in this particular case.

As it turns out, the information provided by the {\it Hipparcos\/}
data is complementary to that given by the ground-based
interferometry, and as we show below the combination of the two allows
the individual masses of \phiher\ to be determined independently of
any assumptions, and independently also of any spectroscopic
information.  Since some of the orbital elements reported by
\cite{Zavala:06} are much improved compared to \cite{Aikman:76}, a
reanalysis of the {\it Hipparcos\/} data with those constraints would
therefore seem to be in order. Better still, the {\it Hipparcos\/}
data can be combined directly with the NPOI observations in a
simultaneous least-squares solution yielding all the elements at
once. This is the path we follow in the next section. The intermediate
data from {\it Hipparcos\/} are available in the form of
one-dimensional `abscissa residuals', which are the residuals from the
standard 5-parameter solutions published in the Catalogue
\citep{ESA:97}.  By extending the 5-parameter model to include orbital
motion, as done by the {\it Hipparcos\/} team, these observations can
be used to strengthen the combined solution.  The nominal errors of
these measurements have a median of 1.85 mas.

\section{Combined orbit and stellar masses}
\label{sec:orbit}

Long-baseline interferometry provides information on the relative
orbit of the binary (with semimajor axis $a$) as well as the
brightness difference, whereas the {\it Hipparcos\/} observations
refer to the motion of the center of light of the binary relative to
the barycenter (with semimajor axis $a_{\rm phot}$), on an absolute
frame of reference. The two kinds of measurements are redundant to
some extent because they both constrain the elements $P$, $e$, $i$,
$\omega_{\rm A}$, $\Omega$, and $T$. Their simultaneous use in a
global fit is therefore expected to lead to a more robust solution.
The use of the {\it Hipparcos\/} measurements introduces several other
parameters that must be solved for at the same time, including
corrections to the catalog values of the position of the barycenter
($\Delta\alpha^*$, $\Delta\delta$) at the mean reference epoch of
1991.25, and corrections to the proper motion components
($\Delta\mu_{\alpha}^*$, $\Delta\mu_{\delta}$) and to the parallax
($\Delta\pi_{\rm Hip}$)\footnote{Following the practice in the {\it
Hipparcos\/} Catalogue we define $\Delta\alpha^* \equiv \Delta\alpha
\cos\delta$ and $\Delta\mu_{\alpha}^* \equiv \Delta\mu_{\alpha}
\cos\delta$.}.  The formalism for modeling the abscissa residuals
follows closely that described by \cite{vanLeeuwen:98},
\cite{Pourbaix:00}, and \cite{Jancart:05}, including the correlations
between measurements from the two independent data reduction consortia
that processed the original {\it Hipparcos\/} observations
\citep{ESA:97}.  The details are reviewed in the Appendix.  With the
inclusion of the semimajor axes $a$ and $a_{\rm phot}$ there are 13
unknowns altogether. We solve for them simultaneously using standard
non-linear least-squares techniques \citep[][p.\ 650]{Press:92}.  In
the course of the iterations we noticed that one of the NPOI
observations (taken on 2004 July 21) gave an unusually large residual
in both angular separation ($+0.86$ mas, or $3.2\sigma$) and position
angle ($-1\fdg16$, or $4.1\sigma$). The same observation stands out in
the solution by \cite{Zavala:06} with similar residuals, and is one of
several measurements taken at nearly the same orbital phase, so is not
particularly critical. We have elected to exclude it from the
solution.  The relative weights between the NPOI and {\it Hipparcos\/}
observations were assigned according to their individual errors. Since
internal errors are not always realistic, we adjusted them by applying
a scale factor in such a way as to achieve a reduced $\chi^2$ value
near unity separately for each type of observation.  This was done by
iterations. A similar procedure was followed by \cite{Zavala:06}. This
resulted in scale factors of 0.34 and 0.24 for the NPOI angular
separations and position angles, and 0.84 for the {\it Hipparcos\/}
measurements.

The results of this fit are given in Table~\ref{tab:elements} under
the heading NPOI+{\it Hipparcos}. For reference we include also the
spectroscopic solution reported by \cite{Aikman:76}, the fit by
\cite{Zavala:06}, and the constrained solution by the {\it
Hipparcos\/} team. The position angle of the node given by
\cite{Zavala:06} requires a change of quadrant to conform to the usual
convention that the ascending node ($\Omega$) corresponds to the node
at which the secondary is receding from the observer.  The orbital
elements from our fit are generally seen to be consistent with those
of \cite{Zavala:06} (whose solution combines interferometry with
radial velocities), but with considerably smaller uncertainties.

\begin{figure} 
\vskip 0.6in
\epsscale{1.0} 
\plotone{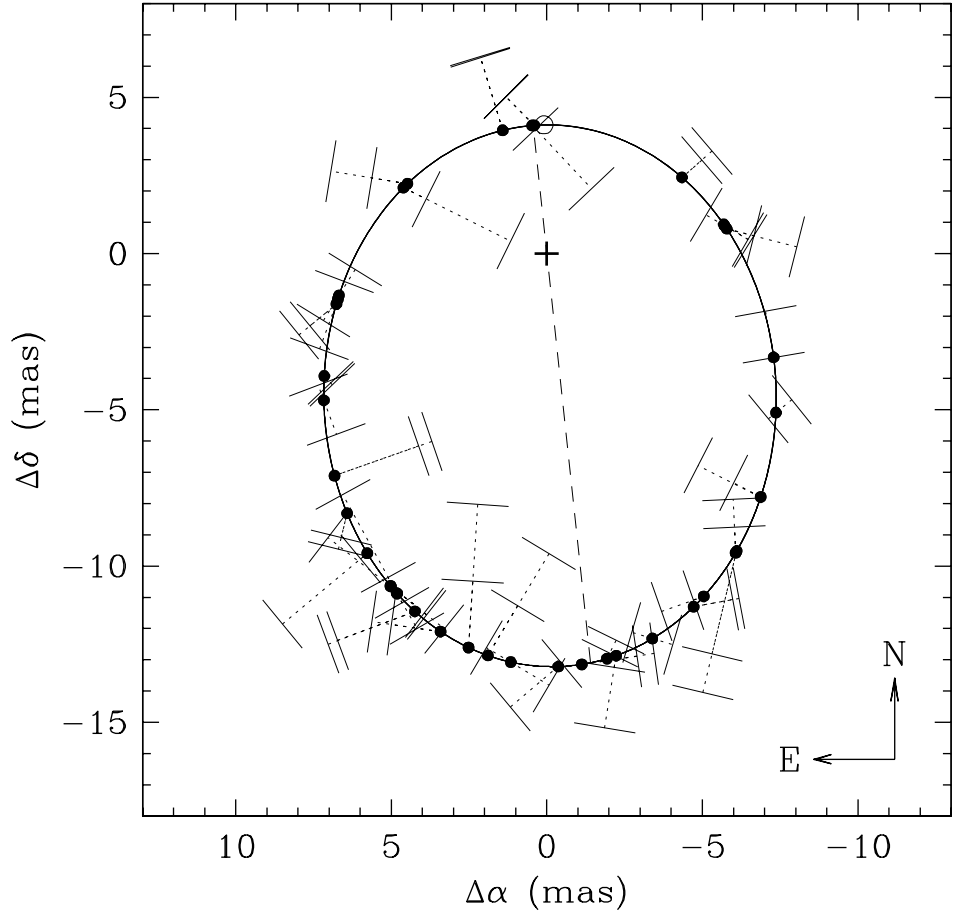}
\vskip 0.0in 

\figcaption[]{Motion of the photocenter of \phiher\ relative to the
center of mass of the binary (indicated by the plus sign) as seen by
{\it Hipparcos}. The solid curve is the computed orbit from our
solution combining {\it Hipparcos\/} and NPOI observations
(Table~\ref{tab:elements}). The one-dimensional abscissa residuals are
shown schematically with a filled circle at the predicted location,
dotted lines representing the scanning direction of the satellite, and
short perpendicular line segments indicating the undetermined location
of the measurement on that line (see text). The length of the dotted
lines represents the magnitude of the $O\!-\!C$ residual from the computed
location.  One measurement with a larger residual was omitted for
clarity. Also indicated on the plot are the location of periastron
(open circle near the top) and the line of nodes (dashed). Motion is
direct (counterclockwise).\label{fig:hiporbit}}

\end{figure}

\begin{figure}
\vskip 0.6in
\epsscale{1.0}
\plotone{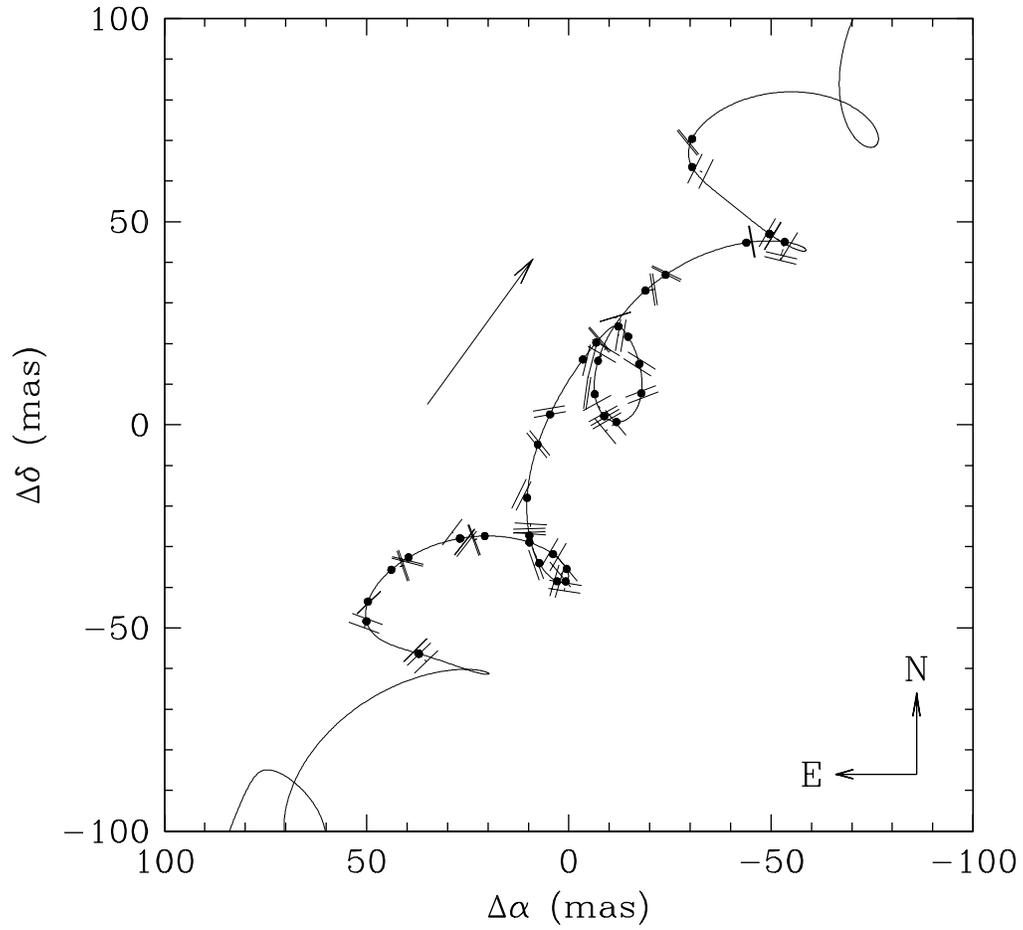}
\vskip 0.0in \figcaption[]{Path of the center of light of \phiher\ on
the plane of the sky, along with the {\it Hipparcos\/} observations
(abscissa residuals) represented as in Figure~\ref{fig:hiporbit}. The
irregular motion is the result of the combined effects of parallax,
proper motion, and orbital motion according to the global solution
described in the text.  The arrow indicates the direction and
magnitude of the annual proper motion.\label{fig:hippath}}
\end{figure}

The projection of the photocentric orbit on the plane of the sky along
with a schematic representation the {\it Hipparcos\/} measurements is
seen in Figure~\ref{fig:hiporbit}, where the axes are parallel to the
right ascension and declination directions.  Because these
measurements are one-dimensional in nature, their exact location on
the plane of the sky cannot be shown graphically. The filled circles
represent the predicted location on the computed orbit.  The dotted
lines connected to each filled circle indicate the scanning direction
of the {\it Hipparcos\/} satellite for each measurement, and show
which side of the orbit the residual is on. The length of each dotted
line represents the magnitude of the $O\!-\!C$ residual.\footnote{The
``$O\!-\!C$ residuals'' are not to be confused with the ``abscissa
residuals'', which we refer to loosely here as {\it Hipparcos\/}
``observations'' or ``measurements''. The abscissa residuals are in
fact residuals from the standard 5-parameter fit reported in the {\it
Hipparcos\/} Catalogue, as stated earlier, whereas the $O\!-\!C$
residuals (or simply ``residuals'') are the difference between the
abscissa residuals and the computed position of the star from a model
that incorporates orbital elements.} The short line segments at the
end of and perpendicular to the dotted lines indicate the direction
along which the actual observation lies, although the precise location
is undetermined. Occasionally more than one measurement was taken
along the same scanning direction, in which case two or more short
line segments appear on the same dotted lines. The orbit is
counterclockwise (direct). The path of \phiher\ on the plane of the
sky as seen by {\it Hipparcos\/} is shown in Figure~\ref{fig:hippath}.
The contorted pattern results from the combination of annual proper
motion (indicated by the arrow), parallactic motion, and orbital
motion.

In addition to providing the parallax, the key contribution of the
{\it Hipparcos\/} observations regarding the masses of \phiher\ is
that they allow the measurement of the semimajor axis of the
photocenter. This parameter can be expressed in terms of the relative
semimajor axis \citep[e.g.,][]{vandeKamp:67} as $a_{\rm phot} =
a(B-\beta)$, where $B$ is the mass fraction $M_{\rm B}/(M_{\rm
A}+M_{\rm B})$ and $\beta$ is the light fraction $\ell_{\rm
B}/(\ell_{\rm A}+\ell_{\rm B})$. This can be put in a more convenient
form as
\begin{equation}
\label{eq:eq1}
a_{\rm phot} = a \left({q\over 1+q} - {1\over 1+10^{0.4 \Delta m}}\right),
\end{equation}
where $q \equiv M_{\rm B}/M_{\rm A}$ and $\Delta m$ is the magnitude
difference in the {\it Hipparcos\/} passband ($H_p$).  Therefore, if
the magnitude difference is known, the combination of both kinds of
astrometric measurements allows one to solve for the mass ratio
$q$. Since the total mass can also be determined from Kepler's Third
Law given the parallax ($\pi_{\rm Hip}$), the individual masses in
units of the solar mass follow immediately:
\begin{eqnarray}
M_{\rm A} & = & {a^3\over \pi_{\rm Hip}^3 P^2}\left(1-{a_{\rm phot}\over
a} - {1\over 1+10^{0.4 \Delta m}}\right) \\
M_{\rm B} & = & {a^3\over \pi_{\rm Hip}^3 P^2}\left({a_{\rm phot}\over
a} + {1\over 1+10^{0.4 \Delta m}}\right).
\end{eqnarray}
In the above expressions all angular quantities are expressed in the
same units, and the period is in units of sidereal years.  The
interferometric measurement of $\Delta m$ by \cite{Zavala:06} is
therefore of crucial importance here, and we have made use of it as an
external constraint (along with its uncertainty) in order to solve for
the masses. Strictly speaking, however, the value entered in the
equations above must correspond to the same passband as the {\it
Hipparcos\/} observations ($H_p$), whereas the relevant $\Delta m$
reported by \cite{Zavala:06} corresponds to the passband of one of the
spectral channels of the NPOI centered at 5500~\AA. We have assumed
that the latter corresponds closely to the visual band, and we have
applied a small correction to transform it to the $H_p$ band using the
relations by \cite{Harmanec:98}. This transformation depends on the
individual $U\!-\!B$ and $B\!-\!V$ colors of the components, which
have not been measured directly. But as described in
\S\ref{sec:models}, they can be estimated using stellar evolution
models (see Table~\ref{tab:colors} below).  The result is then $\Delta
H_p = 2.669 \pm 0.051$, where the uncertainty includes the
contribution from the scatter of the transformation.

The individual masses we obtain for the components of \phiher\ are
reported in Table~\ref{tab:elements}, along with other quantities
derived from the orbital elements, such as the velocity amplitudes
($K_{\rm A}$, $K_{\rm B}$) predicted for the primary and
secondary. The uncertainties in the derived quantities account for the
correlations between the various elements. The total mass is
essentially the same as derived by \cite{Zavala:06}, but with a
significantly reduced error due mostly to the improved
parallax. However, our individual masses are different than the
nominal values they inferred, and are much closer to those expected
for the spectral types estimated by them (see below).  We emphasize
that our mass determinations in this section are purely astrometric,
since we have not made any use of the spectroscopy in our fit.
Several of the orbital elements are considerably more precise in our
solution than in the work of \cite{Zavala:06}, including the semimajor
axis $a$ that in principle depends only on the NPOI observations,
which are common to both fits. This is most likely due to two factors:
our rejection of one of the NPOI observations giving large residuals,
and the fact that we did not use the radial-velocity observations of
\cite{Aikman:76}, which appear to show some inconsistency with the
astrometry and may be biasing the solution of \cite{Zavala:06} causing
larger errors. We investigate this issue in more detail in the next
section. Our error in the period is also smaller despite the shorter
time span of the observations considered in our fit (15.5 yr) compared
to theirs (40.2). We attribute this to the same reasons.

\section{The radial velocities of \phiher}
\label{sec:rvs}

The orbital solution of \cite{Zavala:06} is a combined fit of their
NPOI data with the radial velocities from \cite{Aikman:76}. Although
this resulted in orbital elements with significantly smaller errors,
they instead elected to use the mass function as computed originally
by \cite{Aikman:76}, and with their total system mass of
4.7~M$_{\sun}$ they inferred individual masses of 3.6~M$_{\sun}$ for
the primary and 1.1~M$_{\sun}$ for the secondary. They correctly
pointed out that these values seem anomalous for the spectral types
(which they estimated as B8V and A8V), the primary mass being too
large and the secondary too small. Since the secondary mass is
proportional to $P^{1/3} M_{\rm tot}^{2/3} \sqrt{1-e^2} K_{\rm A}/\sin
i$, the two main factors that contribute to the discrepancy appear to
be the larger inclination angle of \cite{Zavala:06} compared to ours,
and the smaller velocity semiamplitude. Given that the difference in
the inclination angle is not sufficient to explain the problem, we
conclude that the $K_{\rm A}$ value from \cite{Aikman:76} is probably
too small.

\begin{figure}
\vskip 0.6in
\epsscale{1.0}
\plotone{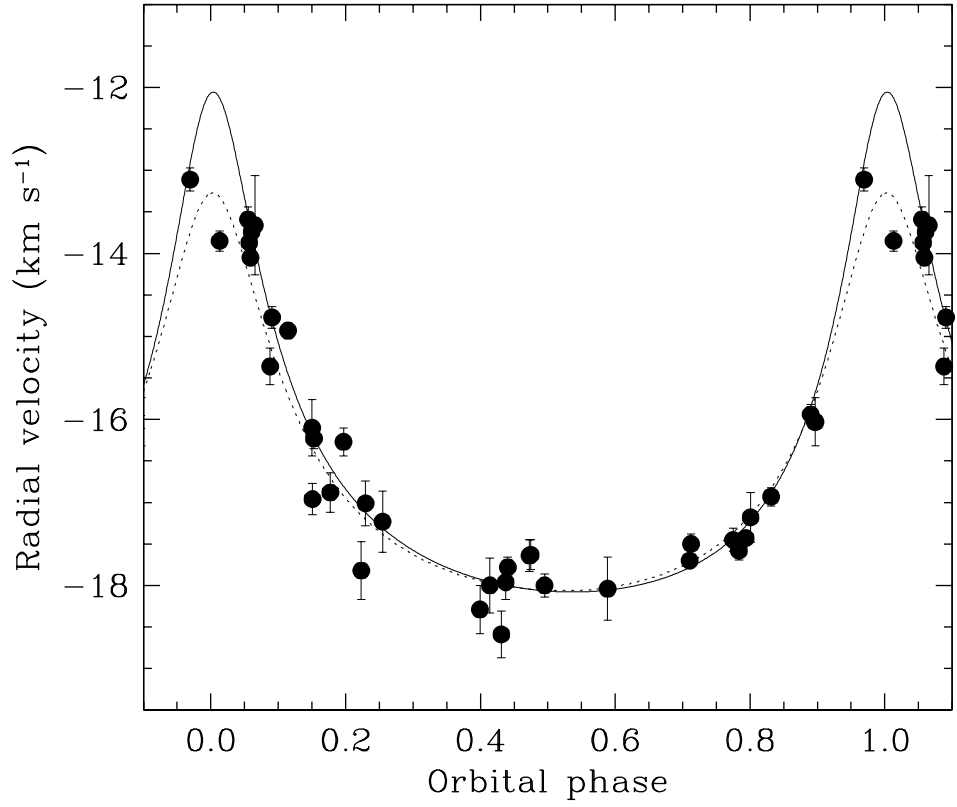}
\vskip -0.5in \figcaption[]{Radial velocity measurements by
\cite{Aikman:76} compared with the predicted velocity curve from our
astrometric orbital solution in \S\ref{sec:orbit} (NPOI+{\it
Hipparcos\/}, solid line), in which the velocities were not used.
Phase 0.0 corresponds to periastron. The dotted line represents the
original spectroscopic solution by
\cite{Aikman:76}.\label{fig:rvcheck}}
\end{figure}

In Figure~\ref{fig:rvcheck} we compare the original radial velocities
by \cite{Aikman:76} with our astrome\-tric solution from the previous
section (indicated with the solid line), in which the center-of-mass
velocity has been adjusted by eye to fit the velocities.  It is seen
that the agreement is very good with the exception of one measurement
obtained precisely at the descending node near phase 0.0, which shows
a residual $\sim$5 times larger than the scatter of the remaining
points. This measurement happens to carry the highest weight for
establishing the semiamplitude $K_{\rm A}$, and we speculate that it
may be biased and may be pulling the amplitude toward lower values.
For reference we show also the original solution by \cite{Aikman:76}
(dotted line), in which the amplitude is lower and that measurement
does not particularly stand out as an outlier. A new spectroscopic
solution without this observation gives a larger semiamplitude
($K_{\rm A} = 2.68 \pm 0.20$~\kms), as expected, although it is also a
bit more uncertain because of the lack of constraint at the
maximum. We note that several of the other elements in this trial
solution are also closer to our results in \S\ref{sec:orbit},
particularly the eccentricity ($e = 0.516 \pm 0.036$) and the
longitude of periastron ($\omega_{\rm A} = 355\fdg6 \pm 3\fdg5$),
which is suggestive.

A search for other velocities of \phiher\ in the literature revealed a
set of measurements of the primary star obtained by \cite{Adelman:01}
between 1989 and 1998 that appears to be of good quality, and that was
not used in the work of \cite{Zavala:06}. Examination of those
data\footnote{We point out that Table~1 of \cite{Adelman:01}
containing the velocities for \phiher\ has the wrong star name in the
title: $\upsilon$~Her instead of \phiher.} shows that unfortunately
they too lack coverage near the all-important descending node, and in
addition there are two outliers\footnote{One, on HJD
$2,\!449,\!134.966$, has a velocity of $-22.0$~\kms\ that is much
lower than any of the others and is clearly erroneous, while the
other, on HJD $2,\!448,\!705.595$, has a velocity nearly 1.5~\kms\
($>$ 3$\sigma$) higher than three other measurements at a similar
phase that show very good interagreement.}. Nevertheless, we believe
them to be potentially useful so we have considered them along with
those of \cite{Aikman:76} in a new combined orbital solution described
next. The velocities of the secondary measured by \cite{Zavala:06}, on
the other hand, are of insufficient quality to be of use and show very
poor agreement with our astrometric solution.

\section{New combined solution and revised masses}
\label{sec:neworbit}

The inclusion of radial velocities in the fit along with the
astrometry introduces a redundancy from the fact that the velocity
semiamplitude is constrained by both kinds of measurements.  In
addition to the 13 adjustable quantities considered previously, we add
the center-of-mass velocity ($\gamma$) and a velocity offset ($\Delta
RV$) to account for a possible difference in the zero points of the
two radial velocity data sets. Furthermore, since the velocities
provide an indirect constraint on $\Delta m$ through eq.[\ref{eq:eq1}]
(see below), we now consider $\Delta m$ also as an adjustable quantity
instead of as a fixed constraint, and the value of the magnitude
difference from interferometry (properly corrected to the $H_p$
passband) is regarded as a measurement with its corresponding
uncertainty. The scale factor for the \cite{Aikman:76} internal errors
was found to be 1.77, and the \cite{Adelman:01} velocities were
assigned uncertainties of 0.47~\kms\ so that their reduced $\chi^2$ is
near unity.

\begin{figure}
\vskip 0.6in
\epsscale{0.9}
\plotone{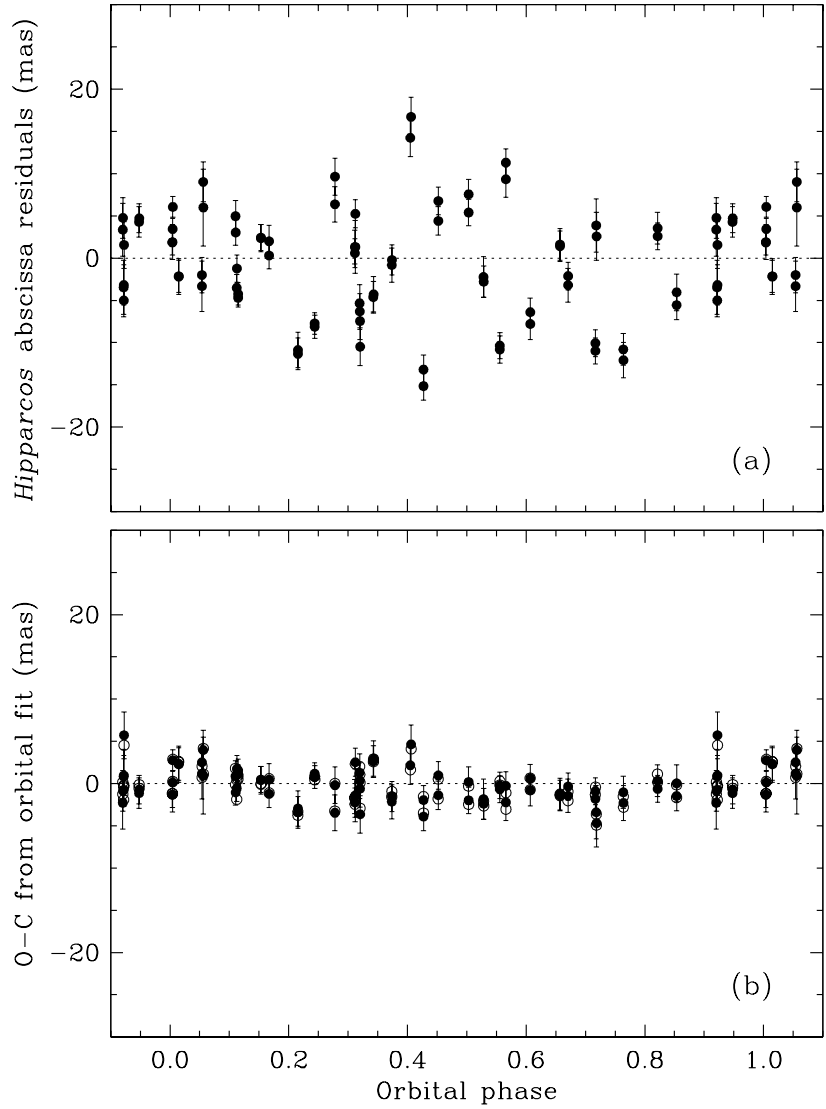}
\vskip -0.5in 

\figcaption[]{(a) Abscissa residuals of \phiher\ from the {\it
Hipparcos\/} mission as a function of orbital phase. These are the
residuals from the standard 5-parameter solution as published in the
Catalogue \citep{ESA:97}. The large scatter in excess of the internal
errors is an indication that there is unmodeled motion in the
system. (b) Filled circles represent the $O\!-\!C$ residuals of the {\it
Hipparcos\/} measurements from our orbital fit that combines
astrometry and radial velocities. The scatter is now much smaller
because the binary motion has been accounted for. The open circles
(without the error bars, to avoid clutter) show the $O\!-\!C$ residuals
resulting from the orbital fit performed by the {\it Hipparcos\/}
team, and are seen to be very similar to ours. \label{fig:hipresid}}

\end{figure}

\begin{figure}
\vskip 0.6in
\epsscale{1.0}
\plotone{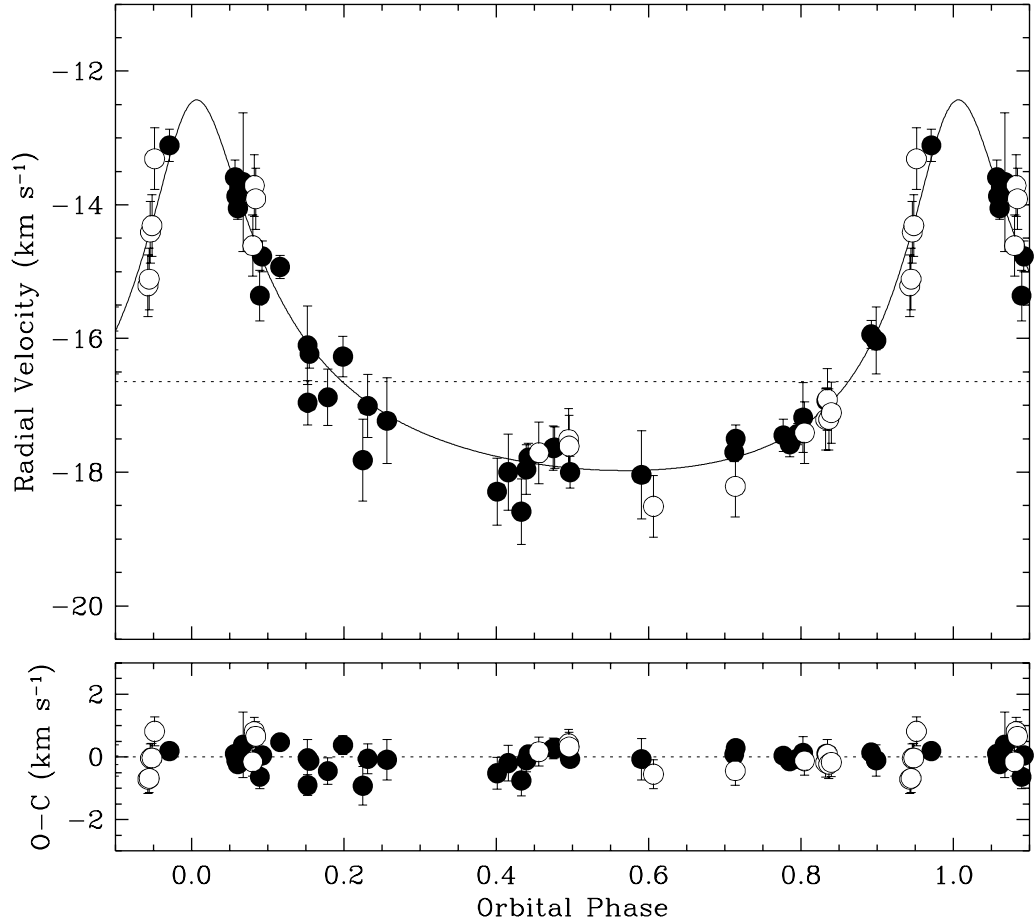}
\vskip -0.6in \figcaption[]{Radial velocity measurements by
\cite{Aikman:76} (filled circles) and \cite{Adelman:01} (open circles)
along with the computed curve from our combined
astrometric-spectroscopic solution. The data by \cite{Adelman:01} have
been adjusted in this figure for the offset $\Delta RV$ found in our
fit (see Table~\ref{tab:elements}). The center-of-mass velocity is
indicated with the dotted line. Residuals are shown at the
bottom.\label{fig:rvorbit}}
\end{figure}

The results of this 16-parameter global solution are listed in the
last column of Table~\ref{tab:elements}. The changes compared to the
previous fit are generally well within the errors, a sign that the
spectroscopic observations we have added are consistent with the
astrometry. The \emph{rms} residuals for an observation of unit weight
(NPOI $\rho$ and $\theta$, radial velocities, {\it Hipparcos\/}) are,
respectively, $\epsilon_{\rho} = 0.13$~mas, $\epsilon_{\theta} =
0\fdg14$, $\epsilon_{\rm RV1} = 0.25$~\kms, $\epsilon_{\rm RV2} =
0.47$~\kms, and $\epsilon_{\rm Hip} = 1.8$~mas.  The individual {\it
Hipparcos\/} measurements and $O\!-\!C$ residuals are listed in
Table~\ref{tab:hipparcos}. In Figure~\ref{fig:hipresid}b we show these
residuals as a function of orbital phase. The $O\!-\!C$ residuals from
the original orbital solution by the {\it Hipparcos\/} team are
overplotted in the same figure (open circles), and the two
distributions are seen to be essentially the same. For reference,
Figure~\ref{fig:hipresid}a shows the {\it Hipparcos\/} measurements on
the same scale \emph{before} accounting for orbital motion, i.e., the
abscissa residuals from the standard 5-parameter solution. The
improvement brought about by the orbital fit is obvious. Minor
systematics at the level of a few milli-arc seconds remain in the
$O\!-\!C$ residuals of Figure~\ref{fig:hipresid}b, for example near
phase 0.7, which we believe may be reflecting the limitations of this
data set.  The residuals from the NPOI observations are quite similar
to those reported by \cite{Zavala:06} and are not repeated here.  The
velocity residuals are given in Table~\ref{tab:aikman} and
Table~\ref{tab:adelman}, along with the original measurements;
observations that were rejected as described above have their
residuals indicated in parentheses.  These data are compared
graphically in Figure~\ref{fig:rvorbit} against the computed velocity
curve from our fit. As an interesting test on the indirect constraint
provided by the velocities on the magnitude difference through other
elements, we carried out another solution in which the magnitude
difference from interferometry is ignored. This fit gives $\Delta H_p
= 3.00 \pm 0.64$~mag, which is not far from the measured value of
$2.669 \pm 0.051$~mag.

\section{Comparison with stellar evolution models}
\label{sec:models}

\begin{figure}
\vskip 0.6in
\epsscale{1.0}
\plotone{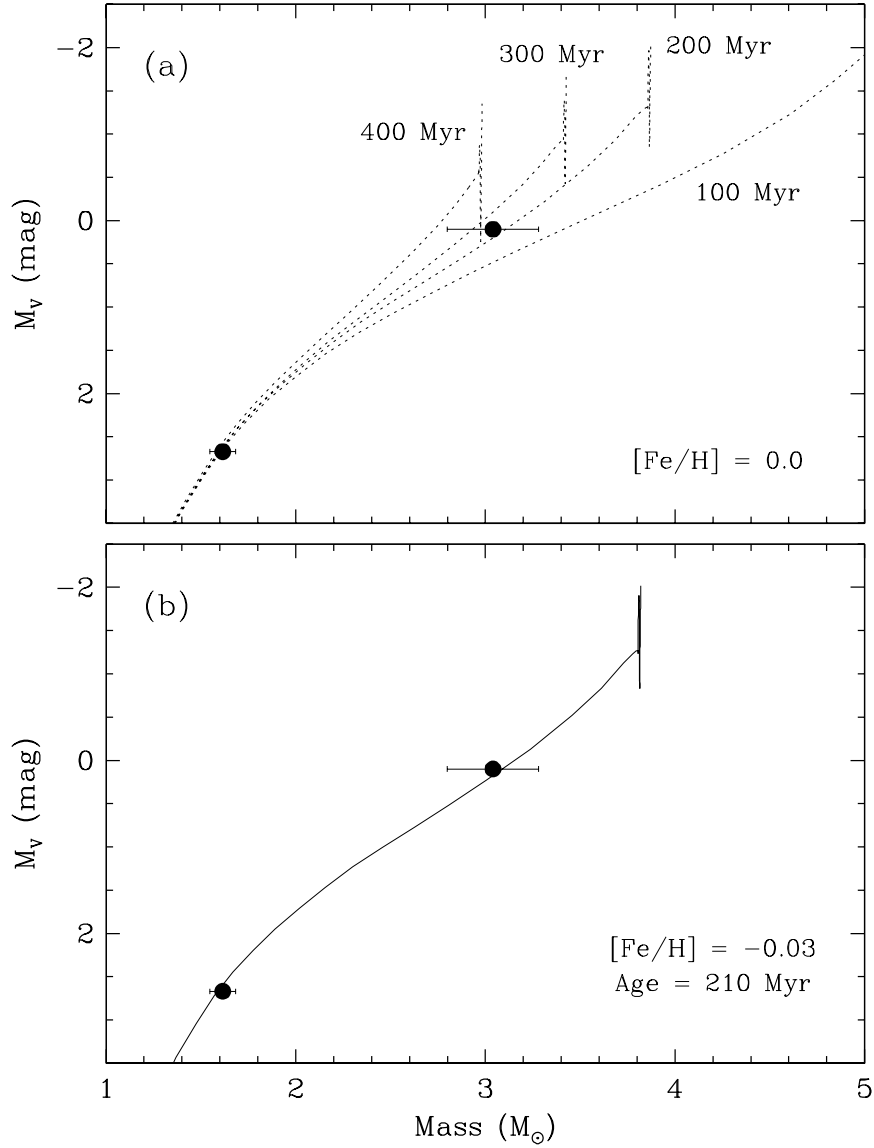}
\vskip -0.8in 

\figcaption[]{Mass-$M_V$ diagram for \phiher. (a) Model isochrones
from the Yonsei-Yale series by \cite{Yi:01} \citep[see
also][]{Demarque:04} for a range of ages, as labeled.  Solar
composition has been assumed. (b) Best-fitting model that reproduces
all six measured properties within their uncertainties (see text). The
individual masses and absolute visual magnitudes as well as the
combined $U\!-\!B$ and $B\!-\!V$ colors are simultaneously matched to better
than 0.4$\sigma$.\label{fig:mlr}}
\end{figure}

As mentioned earlier the individual masses derived here for the
components of \phiher\ are in good agreement with those expected for
the spectral types of the stars. A more stringent comparison may be
made against current stellar evolution models, for example in the
mass-luminosity (or mass-absolute magnitude) plane. The absolute
magnitudes of the stars in the visual band follow from the system
magnitude \citep[$V = 4.220 \pm 0.025$;][]{Mermilliod:94}, the
magnitude difference determined interferometrically ($\Delta V = 2.57
\pm 0.05$), and our revised parallax ($\pi_{\rm Hip}^{\prime} = 14.34
\pm 0.35$~mas), which corresponds to a distance of $69.7 \pm
1.7$~pc. Ignoring extinction we obtain $M_V^{\rm A} = 0.100 \pm 0.059$
and $M_V^{\rm B} = 2.670 \pm 0.074$. In Figure~\ref{fig:mlr}a we show
the measurements against several model isochrones from the Yonsei-Yale
series by \cite{Yi:01} \citep[see also][]{Demarque:04} for ages
between 100~Myr and 400~Myr and solar composition. The agreement is
excellent, and we infer an age of $250 \pm 100$~Myr from this crude
comparison. The assumption of solar composition here is arbitrary, and
in fact a number of detailed analyses of the chemical compositon of
\phiher\ in the literature have generally indicated an iron abundance
above solar. For example, the recent study by \cite{Zavala:06} gave a
value close to [Fe/H] = +0.2. This, however, refers strictly to the
photospheric composition, which is known to be peculiar in \phiher\
and other HgMn stars. In these objects some of the elements are
enhanced by very large factors (up to $10^5$) compared to the Sun,
presumably due to radiatively-driven diffusion and gravitational
settling \citep{Michaud:70}.  Therefore, the interior composition
(which is the relevant quantity for the comparison with models) may be
inaccessible to the observer in this class of stars, except perhaps
through the study of stellar oscillations, should they be detected.

This apparent difficulty has motivated us to turn the argument around
and explore the possibility of inferring the metallicity (as well as
the evolutionary age) from the available observational constraints via
a more systematic comparison with the models, as described below. In
addition to the individual masses and absolute visual magnitudes, we
have considered as constraints the integrated colors of the system,
which are among the easiest quantities to measure accurately.  The
observed $B\!-\!V$ index of \phiher\ as reported by
\cite{Mermilliod:94} is $-0.070 \pm 0.005$ (mean of 46 individual
ground-based measurements). On the other hand the {\it Hipparcos\/}
Catalogue gives the considerably redder value $-0.045 \pm 0.003$,
based on the $B_{\rm T}$ and $V_{\rm T}$ measurements from the {\it
Tycho\/} experiment onboard the satellite along with a conversion to
the Johnson system. These later measurements were superseded by the
re-reduction that resulted in the {\it Tycho-2\/} Catalogue
\citep{Hog:00}, according to which $B_{\rm T} = 4.155 \pm 0.014$ and
$V_{\rm T} = 4.216 \pm 0.009$. Based on these revised values and the
conversion to the Johnson system described in the {\it Hipparcos\/}
Catalogue \citep[][Vol.\ 1, Sect.\ 1.3]{ESA:97} we obtain $B\!-\!V =
-0.055 \pm 0.014$.  This is still redder than the ground-based
average, although the uncertainty is perhaps more realistic so that
the two determinations are now consistent within the errors. The
weighted average, which we adopt in the following, is $\langle
B\!-\!V\rangle = -0.068 \pm 0.008$. The $U\!-\!B$ color of \phiher\ as
reported by \cite{Mermilliod:94} is $U\!-\!B = -0.250 \pm 0.009$, from
the mean of 40 individual ground-based observations.

\begin{figure}
\vskip 0.6in
\epsscale{1.0}
\plotone{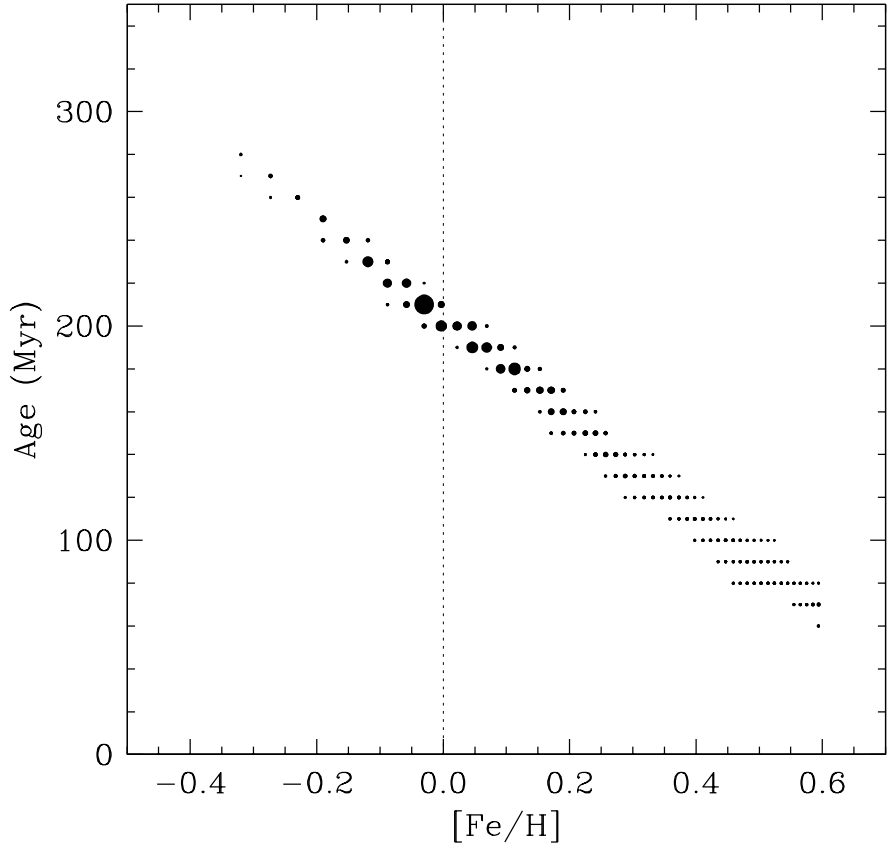}
\vskip -1.8in 

\figcaption[]{Age and metallicity for all model isochrones by
\cite{Yi:01} \citep[see also][]{Demarque:04} that yield a match to the
measured properties of \phiher\ (individual masses and absolute visual
magnitudes, and combined $U\!-\!B$ and $B\!-\!V$ colors) within the
observational errors. The size of the points is an indication of the
quality of the agreement, with larger symbols representing a better
fit to the data (see text). Solar composition is represented with a
dotted line, for reference. The best fit corresponds to [Fe/H] $=
-0.03$ and an age of 210~Myr.\label{fig:agemet}}

\end{figure}

To compare the six measured properties ($M_{\rm A}$, $M_{\rm B}$,
$M_V^{\rm A}$, $M_V^{\rm A}$, $U\!-\!B$, $B\!-\!V$) against
evolutionary models, we computed by interpolation a large grid of
Yonsei-Yale isochrones spanning a wide range of ages (50 to 500~Myr)
and metallicities ($-0.50$ to +0.60 in [Fe/H]). Along each isochrone
we interpolated the colors and magnitudes of stars in a fine grid of
masses over intervals of $\pm$1$\sigma$ centered on our measured
values of $M_{\rm A}$ and $M_{\rm B}$. For all combinations of a
primary and secondary mass taken from these two intervals we
calculated the theoretical integrated colors of the system, and
recorded all combinations of the isochrone age and metallicity that
yielded simultaneous agreement with the measured individual absolute
magnitudes and combined $U\!-\!B$ and $B\!-\!V$ indices, within their
uncertainties.  Figure~\ref{fig:agemet} displays all consistent models
in the age/metallicity plane, where the point size is related to the
goodness of fit as measured by the distance $D^2$ between the model and
the observations in the six-dimensional parameter space,
\begin{eqnarray*}
D^2 & = & \left[\Delta M_{\rm A}\right]^2 + \left[\Delta M_{\rm
B}\right]^2 + \left[\Delta M_V^{\rm A}\right]^2 + \left[\Delta
M_V^{\rm B}\right]^2 + \\
& + & \left[\Delta (U\!-\!B)\right]^2 + \left[\Delta
(B\!-\!V)\right]^2  .
\end{eqnarray*}
Age and metallicity are seen to be highly correlated. The best match
with the observations occurs near the middle of the
distribution\footnote{Although a heavy element abundance very much
higher than the Sun (even beyond the range we explored) would appear
to be allowed by the models (but with lower significance, as seen in
the figure), and would imply considerably younger ages, we believe
those scenarios to be unlikely.} for a composition very near solar
([Fe/H] $= -0.03$) and an age of 210~Myr.  The individual masses
preferred by this model are within 0.4$\sigma$ of the measured values,
the agreement with the individual absolute magnitudes is virtually
perfect, and that with the integrated colors is $\sim$0.1$\sigma$.  In
Table~\ref{tab:colors} we list the predicted properties from this
model along with the measured values. The individual $U\!-\!B$ and
$B\!-\!V$ colors for the components are the ones used in
\S\ref{sec:orbit} to convert the interferometric magnitude difference
$\Delta V$ into $\Delta H_p$.  The best-fit isochrone is shown in
Figure~\ref{fig:mlr}b.  Based on this comparison one may conclude that
the interior composition of \phiher\ is not significantly different
from solar, which is more or less as expected for a young early-type
system such as this. We point out, however, that this relies on the
colors of \phiher\ being normal and also on the colors predicted by
the stellar evolution models being realistic. Theoretical colors are
unlikely to be in error by significant amounts, but we do note that
there are some indications of anomalies in the spectrophotometry of
HgMn stars \citep[see, e.g.,][and references therein]{Adelman:84,
Adelman:03} of a nature similar to those seen in metallic-line A stars
and other chemically peculiar objects, which are thought to be
connected with line blanketing, particularly in the bluer spectral
regions.

\section{Concluding remarks}

The orbital solutions in \S\ref{sec:orbit} and \S\ref{sec:neworbit}
provide an interesting illustration of the usefulness of the {\it
Hipparcos\/} intermediate data as a valuable complement to other
astrometric or spectroscopic observations. The abscissa residuals from
the satellite mission are publicly available for many thousands of
binary stars distributed over the entire sky, and although a number of
studies have appeared over the past few years that do take advantage
of them, this is largely still an underutilized resource.  In the case
of the HgMn star \phiher\ the {\it Hipparcos\/} measurements are key
to establishing the individual masses of the components, the most
fundamental of the stellar properties. We find good agreement between
these masses and the magnitudes and colors of the system and current
stellar evolution models, suggesting the bulk composition of the
object is near solar.  Although the faint secondary has now been
detected, both interferometrically and spectroscopically, an accurate
measurement of its radial velocity over the orbital cycle should help
considerably for reducing the mass uncertainties, which are currently
at the 8\% and 4\% level for the primary and secondary, respectively.
	
\acknowledgements 

We thank the referee for a number of helpful comments on the original
manuscript.  This work was partially supported by NSF grant
AST-0406183 and NASA's MASSIF SIM Key Project (BLF57-04). This
research has made use of the SIMBAD database, operated at CDS,
Strasbourg, France, and of NASA's Astrophysics Data System Abstract
Service.

\appendix

\section{Making use of the {\it Hipparcos\/} intermediate observations
in the orbital solution of \phiher}
\label{appendix}

The intermediate data provided with the {\it Hipparcos\/} catalog are
the ``abscissa residuals'', $\Delta v$, which represent the difference
between the satellite measurements (abscissae) along great circles and
the predicted abscissae computed from the 5 standard astrometric
parameters. The standard parameters are the position of the object
($\alpha_0^*$, $\delta_0$) at the reference epoch $t_0 = 1991.25$, the
proper motion components ($\mu_{\alpha}^*$, $\mu_{\delta}$), and the
parallax ($\pi_{\rm Hip}$). We follow here the notation in the {\it
Hipparcos\/} catalog and define $\alpha_0^* \equiv \alpha_0
\cos\delta$ and $\mu_{\alpha}^* \equiv \mu_{\alpha} \cos\delta$, to
include the projection factors.  The goal of incorporating an orbital
model into the analysis of the {\it Hipparcos\/} data is to reduce the
original abscissa residuals to values below those obtained from the
5-parameter solution by taking into account the orbital motion of the
photocenter.\footnote{Regardless of the actual model used by the {\it
Hipparcos\/} team to obtain the final published solution, the abscissa
residuals provided are always those resulting from the five standard
parameters as listed in the catalog. This allows complete generality
in extending the model beyond those five parameters to include orbital
motion or other motions of arbitrary complexity.} The $\chi^2$
minimization approach for doing this has been described previously by
\cite{vanLeeuwen:98}, \cite{Pourbaix:00}, \cite{Jancart:05}, and
others.  We review the procedure here with additional details to
facilitate its application in other cases.  Following
\cite{Pourbaix:00} the $\chi^2$ sum can be represented quite generally
as $\chi^2 = \mbox{\boldmath $\Xi^t~ V^{-1}~ \Xi$}$, where
\begin{equation} 
\label{eq:chiterm} 
\mbox{\boldmath $\Xi$} = \mbox{\boldmath $\Delta v$} - \sum_{k=1}^M
{\partial \mbox{\boldmath $v$}\over\partial p_k} \Delta p_k
\end{equation}
and \mbox{\boldmath $\Xi^t$} is the transpose of \mbox{\boldmath
$\Xi$}.  In this expression \mbox{\boldmath $\Delta v$} is the array
of N abscissa residuals provided by {\it Hipparcos}, and $\partial
\mbox{\boldmath $v$}/\partial p_k$ is the array of partial derivatives
of the abscissae with respect to the $k$th fitted parameter.  The
number $M$ of parameters fitted to the astrometry in our case is 12:
the 5 standard {\it Hipparcos\/} parameters ($p_1 = \alpha_0^*$, $p_2
= \delta_0$, $p_3 = \mu_{\alpha}^*$, $p_4 = \mu_{\delta}$, $p_5 =
\pi_{\rm Hip}$) and 7 orbital elements ($a_{\rm phot}$, $P$, $e$, $i$,
$\omega_{\rm A}$, $\Omega$, $T$, represented as $p_k$ with
$k=6,\ldots,12$). Here \mbox{\boldmath $V^{-1}$} is the inverse of the
covariance matrix of the observations, containing the abscissa
uncertainties and correlation coefficients \citep[][Vol.\ 3, eqs.\
17-10 and 17-11]{ESA:97}. These are both provided in the {\it
Hipparcos\/} catalog for each observation.  Correlations arise because
the same original satellite data were reduced independently by two
data reduction consortia \citep[FAST and NDAC; see][]{ESA:97}, and the
results from both are typically included in all solutions. To be
explicit,
{\boldmath
\[ V = \left(
\begin{array}{cccc}
V_1 & 0 & \ldots & 0 \\
0   & V_2 & \ddots & \vdots \\
\vdots & \ddots & \ddots & 0 \\
0 & \ldots & 0 & V_n
\end{array}
\right) \]
}
where each subarray {\boldmath $V_j$} corresponds to a pair of
FAST/NDAC measurements (F/N) and is given by
\begin{equation} \mbox{\boldmath $V_j$} = \left(
\begin{array}{cc}
\sigma_{\rm F}^2 & \rho\sigma_{\rm F}\sigma_{\rm N} \\ \rho\sigma_{\rm
F}\sigma_{\rm N} & \sigma_{\rm N}^2
\end{array}
\right)_j \end{equation}
in which $\sigma_{\rm F}$ and $\sigma_{\rm N}$ are the corresponding
uncertainties and $\rho$ is the correlation coefficient.  

The partial derivatives $\partial \mbox{\boldmath $v$}/\partial p_k$
for $k=1$ to 5 in equation (\ref{eq:chiterm}) are given in the {\it
Hipparcos\/} catalog along with the abscissa residuals. The remaining
derivatives can be expressed in terms of the partial derivatives of
\mbox{\boldmath $v$} with respect to $\alpha_0^*$ and
$\delta_0$. These are \citep[][Vol.\ 3, eq.\ 17-15]{ESA:97}
\begin{equation} 
\label{eq:derivative}
{\partial\mbox{\boldmath $v$}\over\partial p_k} = {\partial
\mbox{\boldmath $v$}\over\partial\alpha_0^*}{\partial\xi\over\partial
p_k}+{\partial\mbox{\boldmath
$v$}\over\partial\delta_0}{\partial\eta\over\partial
p_k}~~,~~k=6,\ldots,12,
\end{equation}
in which $\xi$ and $\eta$ are the rectangular coordinates of the
photocenter relative to the center of mass of the binary on the plane
tangent to the sky at ($\alpha_0^*$, $\delta_0$). The general
expressions for these are
\begin{eqnarray*}
\label{eq:xi}
\xi & = & \alpha_0^* + \mu_{\alpha}^*(t-t_0) + P_{\alpha}\pi_{\rm Hip}
+ \Delta X \\
\label{eq:eta}
\eta & = & \delta_0 + \mu_{\delta}(t-t_0) + P_{\delta}\pi_{\rm Hip} +
\Delta Y
\end{eqnarray*}
in which $P_{\alpha}$ and $P_{\delta}$ are the parallactic factors.
Only the last term in each of these equations is relevant in our
case. They represent the right ascension and declination components of
the orbital motion, and are conveniently expressed as $\Delta X = B x
+ G y$ and $\Delta Y = A x + F y$. Here $x$ and $y$ are the
rectangular coordinates in the unit orbit given by $x = \cos E - e$
and $y = \sqrt{1-e^2} \sin E$, with $E$ being the eccentric anomaly.
This angle is related to the period and time of periastron passage
through Kepler's equation, $E - e \sin E = 2\pi(t-T)/P$.  The symbols
$A$, $B$, $F$, and $G$ are the classical Thiele-Innes constants
\citep[see, e.g.,][]{vandeKamp:67}, which depend only on the orbital
elements $a_{\rm phot}$, $i$, $\omega_{\rm A}$, and $\Omega$, and are
given by
\begin{eqnarray}
B & = & -a_{\rm phot}(\cos\omega_{\rm A} \sin\Omega + \sin\omega_{\rm A} \cos\Omega \cos i) \nonumber \\
A & = & -a_{\rm phot}(\cos\omega_{\rm A} \cos\Omega - \sin\omega_{\rm A} \sin\Omega \cos i) \nonumber \\
G & = & -a_{\rm phot}(-\sin\omega_{\rm A} \sin\Omega + \cos\omega_{\rm A} \cos\Omega \cos i) \nonumber \\
F & = & -a_{\rm phot}(-\sin\omega_{\rm A} \cos\Omega - \cos\omega_{\rm A} \sin\Omega \cos i) \nonumber
\end{eqnarray}
The negative sign preceding $a_{\rm phot}$ in these equations reflects
our use for this particular application of the longitude of periastron
for the primary ($\omega_{\rm A}$) instead of that of the secondary
($\omega_{\rm B}$), for consistency with the elements of the
spectroscopic orbit. The customary form of the Thiele-Innes constants
in solving the relative orbit of a visual binary uses $\omega_{\rm
B}$.  Trivially the two angles differ by 180\arcdeg.

As described by \cite{Pourbaix:00}, the nature of the orbital solution
is such that of the seven derivatives in equation
(\ref{eq:derivative}) the only one that needs to be considered
explicitly is the derivative with respect to the semimajor axis,
$\partial \mbox{\boldmath $v$}/\partial p_6 = \partial \mbox{\boldmath
$v$}/\partial a_{\rm phot}$. The expression for \mbox{\boldmath $\Xi$}
in equation (\ref{eq:chiterm}) then reduces to
\begin{displaymath} 
\mbox{\boldmath $\Xi$} = \mbox{\boldmath $\Delta v$} - \sum_{k=1}^5
{\partial \mbox{\boldmath $v$}\over\partial p_k} \Delta p_k -
\left({\partial \mbox{\boldmath $v$}\over\partial\alpha_0^*}
{\partial\xi\over\partial a_{\rm phot}} + {\partial \mbox{\boldmath
$v$}\over\partial\delta_0} {\partial\eta\over\partial a_{\rm
phot}}\right) a_{\rm phot}~.
\end{displaymath}
The remaining six orbital elements ($P$, $e$, $i$, $\omega_{\rm A}$,
$\Omega$, and $T$) do not appear explicitly but are hidden in
$\partial\xi/\partial a_{\rm phot} = \partial(\Delta X)/\partial
a_{\rm phot}$ and $\partial\eta/\partial a_{\rm phot} =
\partial(\Delta Y)/\partial a_{\rm phot}$. Thus, 12 parameters enter
the evaluation of $\chi^2$ and are to be adjusted to seek its minimum:
six are explicit ($a_{\rm phot}$ and $\Delta p_k$, with $k =
1,\ldots,5$), and the remaining six are implicit.

At each iteration towards the $\chi^2$ minimum the array of $O\!-\!C$
residuals \mbox{\boldmath $\Xi$} is properly weighted by accounting
for the error of each abscissa residual and the correlation between
the FAST and NDAC measurements, which typically come in
pairs. Representing one of such pairs by $\mbox{$\mathbf \Xi$}_j =
\left(r_{\rm F}, r_{\rm N}\right)_j$, it is easy to see using the
definition of {\boldmath $V_j$} above that the corresponding $j$th
term $\mbox{$\mathbf \Xi$}^t_j$ {\boldmath $V_j^{-1}$} $\mbox{$\mathbf
\Xi$}_j$ in the $\chi^2$ sum will be
\begin{displaymath}
{1\over 1-\rho^2} \left[ \left({r_{\rm F}\over\sigma_{\rm F}}\right)^2
- {2\rho r_{\rm F} r_{\rm N}\over\sigma_{\rm F}\sigma_{\rm N}} +
\left({r_{\rm N}\over\sigma_{\rm N}}\right)^2 \right]~.
\end{displaymath}
If for some reason a measurement for only one consortium is available
on a certain date (i.e., the observations are not paired for that
particular orbit of the {\it Hipparcos\/} satellite), the correlation
coefficient is zero and the $\chi^2$ term reduces to $(r_{\rm
F}/\sigma_{\rm F})^2$ or $(r_{\rm N}/\sigma_{\rm N})^2$.

\clearpage

\begin{deluxetable}{lccccc}
\tabletypesize{\scriptsize}
\tablecolumns{3}
\tablewidth{0pc}
\tablecaption{Orbital solutions for \phiher.\label{tab:elements}}
\tablehead{\colhead{} & \colhead{\cite{Aikman:76}} & \colhead{\cite{Zavala:06}} & \colhead{} & \colhead{This paper} & \colhead{This paper} \\
\colhead{\hfil~~~~~~~~~~~Parameter~~~~~~~~~~~~} & \colhead{(RVs)} & \colhead{(NPOI+RVs)} & \colhead{{\it Hipparcos}} & \colhead{(NPOI+{\it Hipparcos})} & \colhead{(NPOI+{\it Hipparcos}+RVs)} }
\startdata
\noalign{\vskip -6pt}
\sidehead{Adjusted quantities} \\
\noalign{\vskip -9pt}
~~~~$P$ (days)\dotfill                         &  560.5~$\pm$~1.7\phn\phn     &  564.69~$\pm$~0.13\phn\phn   &  560.5\tablenotemark{a} &  564.783~$\pm$~0.048\phn\phn & 564.834~$\pm$~0.038\phn\phn \\
~~~~$\gamma$ (\kms)\dotfill                    &  $-16.79$~$\pm$~0.06\phs\phn &  $-16.66$~$\pm$~0.05\phs\phn &  \nodata                &  \nodata & $-16.642$~$\pm$~0.045\phn\phs \\
~~~~$\Delta RV$ (\kms)\tablenotemark{b}\dotfill  &  \nodata                   &  \nodata                     &  \nodata                &  \nodata & $-0.41$~$\pm$~0.12\phs \\
~~~~$K_{\rm A}$ (\kms)\dotfill                 &  2.39~$\pm$~0.12             &  2.5                         &  \nodata                &  3.02~$\pm$~0.28\tablenotemark{c} & 2.772~$\pm$~0.073\tablenotemark{c} \\
~~~~$e$\dotfill                                &  0.47~$\pm$~0.03             &  0.522~$\pm$~0.004           &  0.47\tablenotemark{a}  &  0.5250~$\pm$~0.0011 & 0.52614~$\pm$~0.00086 \\
~~~~$\omega_{\rm A}$ (deg)\dotfill             &  357~$\pm$~5\phn\phn         &  351.9~$\pm$~2.7\phn\phn     &  357\tablenotemark{a}   &  355.0~$\pm$~4.4\phn\phn & 350.8~$\pm$~1.4\phn\phn \\
~~~~$T$ (HJD$-$2,400,000)\dotfill              & \phm{\tablenotemark{b}}50053.7~$\pm$~5.5\tablenotemark{d}\phm{2222}  & 50121.8~$\pm$~1.0\phm{2222}  &  \phm{\tablenotemark{e}}50114~$\pm$~16\tablenotemark{d}\phm{222} & 50121.68~$\pm$~0.25\phm{2222} & 50121.43~$\pm$~0.20\phm{2222} \\
~~~~$i$ (deg)\dotfill                          &  \nodata                     &  12.1~$\pm$~2.9\phn          &  36~$\pm$~14            & 9.80~$\pm$~0.77 & 9.10~$\pm$~0.40 \\
~~~~$\Omega_{\rm J2000}$ (deg)\dotfill         &  \nodata                     &  \phm{\tablenotemark{a}}189.1~$\pm$~2.5\tablenotemark{e}\phn\phn & 188~$\pm$~12\phn & 186.2~$\pm$~4.4\phn\phn & 190.4~$\pm$~1.4\phn\phn \\
~~~~$a$ (mas)\dotfill                          &  \nodata                     &  32.1~$\pm$~0.2\phn          &  \nodata & 32.045~$\pm$~0.035\phn & 32.027~$\pm$~0.028\phn \\
~~~~$a_{\rm phot}$ (mas)\dotfill               &  \nodata                     &  \nodata                     &  9.09~$\pm$~0.65 & 8.67~$\pm$~0.39 & 8.57~$\pm$~0.36 \\
~~~~$\Delta\alpha^*$ (mas)\dotfill             &  \nodata                     &  \nodata                     &  \nodata & +0.45~$\pm$~0.32\phs & +0.44~$\pm$~0.32\phs \\
~~~~$\Delta\delta$ (mas)\dotfill               &  \nodata                     &  \nodata                     &  \nodata & $-0.58$~$\pm$~0.43\phs & $-0.52$~$\pm$~0.42\phs \\
~~~~$\Delta\mu_{\alpha}^*$ (mas~yr$^{-1}$)\dotfill   &  \nodata               &  \nodata                     &  \nodata & $-0.17$~$\pm$~0.32\phs & $-0.15$~$\pm$~0.32\phs \\
~~~~$\Delta\mu_{\delta}$ (mas~yr$^{-1}$)\dotfill     &  \nodata               &  \nodata                     &  \nodata & +0.01~$\pm$~0.35\phs & +0.03~$\pm$~0.34\phs \\
~~~~$\Delta\pi_{\rm Hip}$ (mas)\dotfill        &  \nodata                     &  \nodata                     &  \nodata & +0.02~$\pm$~0.36\phs & +0.07~$\pm$~0.35\phs \\
~~~~$\Delta m$ (mag)\dotfill                   &  \nodata                     &  2.57~$\pm$~0.05\tablenotemark{f}  &  \nodata & 2.669~$\pm$~0.051\tablenotemark{g} & 2.672~$\pm$~0.052\tablenotemark{h} \\
\sidehead{Derived quantities} \\						                                         
\noalign{\vskip -9pt}										                                         
~~~~$K_{\rm B}$ (\kms)\dotfill                 &  \nodata                     &  8.1                         &  \nodata & 5.62~$\pm$~0.49 & 5.23~$\pm$~0.29 \\
~~~~$M_{\rm A}$ (M$_{\sun}$)\dotfill           &  \nodata                     &  3.6                         &  \nodata & 3.07~$\pm$~0.25 & 3.05~$\pm$~0.24 \\
~~~~$M_{\rm B}$ (M$_{\sun}$)\dotfill           &  \nodata                     &  1.1                         &  \nodata & 1.647~$\pm$~0.075 & 1.614~$\pm$~0.066 \\
~~~~$q\equiv M_{\rm B}/M_{\rm A}$\dotfill      &  \nodata                     &  0.31                        &  \nodata & 0.537~$\pm$~0.031 & 0.530~$\pm$~0.027 \\
~~~~$M_{\rm A}+M_{\rm B}$ (M$_{\sun}$)\dotfill &  \nodata                     &  4.7~$\pm$~0.6               &  \nodata & 4.71~$\pm$~0.37 & 4.66~$\pm$~0.34 \\
~~~~$\mu_{\alpha}^*$ (mas~yr$^{-1}$)\dotfill   &  \nodata                     &  \nodata                     & $-25.98$~$\pm$~0.45\phs\phn & $-26.15$~$\pm$~0.32\phn\phs & $-26.13$~$\pm$~0.32\phn\phs \\
~~~~$\mu_{\delta}$ (mas~yr$^{-1}$)\dotfill     &  \nodata                     &  \nodata                     & +35.86~$\pm$~0.48\phs\phn & +35.87~$\pm$~0.35\phn\phs  & +35.89~$\pm$~0.34\phn\phs \\
~~~~$\pi_{\rm Hip}$ (mas)\dotfill              &  \nodata                     &  \nodata                     & 14.27~$\pm$~0.52\phn & 14.29~$\pm$~0.36\phn & 14.34~$\pm$~0.35\phn \\
\sidehead{Other quantities pertaining to the fit} \\						                                         
\noalign{\vskip -9pt}										                                         
~~~~$N_{\rm RV}$\dotfill                           & 37                           & 37                           & \nodata              & \nodata              & 36 + 18 \\
~~~~$N_{\rm NPOI}$ ($\rho$, $\theta$)\dotfill      & \nodata                      & 25 + 25                      & \nodata              & 24 + 24              & 24 + 24 \\
~~~~$N_{\rm Hip}$\dotfill                          & \nodata                      & \nodata                      & 76                   & 76                   & 76      \\
~~~~Total time span (yr)\dotfill                   & 8.4                          & 40.2                         & 3.3                  & 15.5                 & 40.2    \\
\enddata
\tablenotetext{a}{Value adopted from the solution by \cite{Aikman:76}, and held fixed.}
\tablenotetext{b}{Systematic radial velocity offset in the sense $\langle$\cite{Adelman:01} \emph{minus} \cite{Aikman:76}$\rangle$.}
\tablenotetext{c}{Parameter derived from other elements in this solution, as opposed to being adjusted.}
\tablenotetext{d}{Shifted forward by an integer number of cycles from the published epoch in order to match other solutions.}
\tablenotetext{e}{Quadrant reversed from published value.}
\tablenotetext{f}{Corresponds to a wavelength of 5500~\AA, and was derived from the interferometric visibilities separately from the orbital solution by holding the orbital elements fixed.}
\tablenotetext{g}{Converted to the {\it Hipparcos\/} passband ($H_p$) and held fixed in the solution, as an external constraint.}
\tablenotetext{h}{Corresponds to the {\it Hipparcos\/} passband ($H_p$).}
\end{deluxetable}

\clearpage

\LongTables
\begin{deluxetable}{lccccc}
\tabletypesize{\scriptsize}
\tablewidth{0pc}
\tablecaption{{\it Hipparcos\/} measurements of \phiher\ and corresponding $O\!-\!C$ residuals.\label{tab:hipparcos}}
\tablehead{\colhead{HJD} & \colhead{} & \colhead{$v$\tablenotemark{a}} & \colhead{$\sigma_{v}$\tablenotemark{b}} & \colhead{$O\!-\!C$} & \colhead{} \\
\colhead{\hbox{~~(2,400,000$+$)~~}} & \colhead{Julian Year} & \colhead{(mas)} & \colhead{(mas)} & \colhead{(mas)} & \colhead{Phase}}
\startdata
 47864.4610\dotfill &   1989.9232 &  \phn$+$1.91 &  2.08 &   $-$1.30 &  0.0042 \\
 47864.3440\dotfill &   1989.9229 &  \phn$+$1.86 &  1.50 &   $-$1.32 &  0.0040 \\
 47864.6710\dotfill &   1989.9238 &  \phn$+$3.44 &  1.29 &   $+$0.14 &  0.0045 \\
 47864.7937\dotfill &   1989.9241 &  \phn$+$6.07 &  1.23 &   $+$2.76 &  0.0048 \\
 47925.7283\dotfill &   1990.0910 &  \phn$-$1.23 &  1.60 &   $+$1.74 &  0.1126 \\
 47925.5691\dotfill &   1990.0905 &  \phn$-$3.51 &  1.55 &   $-$0.56 &  0.1124 \\
 47948.6166\dotfill &   1990.1536 &  \phn$+$2.36 &  1.62 &   $+$0.40 &  0.1532 \\
 47948.6800\dotfill &   1990.1538 &  \phn$+$2.44 &  1.55 &   $+$0.48 &  0.1533 \\
 47983.8125\dotfill &   1990.2500 &  $-$11.33 &  1.87 &   $-$3.39 &  0.2155 \\
 47983.8512\dotfill &   1990.2501 &  $-$10.88 &  2.09 &   $-$2.95 &  0.2155 \\
 47999.8375\dotfill &   1990.2939 &  \phn$-$8.12 &  1.35 &   $+$0.79 &  0.2439 \\
 47999.7452\dotfill &   1990.2936 &  \phn$-$7.73 &  1.29 &   $+$1.18 &  0.2437 \\
 48038.0982\dotfill &   1990.3986 &  \phn$+$0.59 &  1.73 &   $-$2.24 &  0.3116 \\
 48037.9790\dotfill &   1990.3983 &  \phn$+$1.31 &  2.01 &   $-$1.52 &  0.3114 \\
 48038.3088\dotfill &   1990.3992 &  \phn$+$1.32 &  3.12 &   $-$1.39 &  0.3120 \\
 48038.4336\dotfill &   1990.3995 &  \phn$+$5.24 &  1.66 &   $+$2.52 &  0.3122 \\
 48055.6215\dotfill &   1990.4466 &  \phn$-$4.62 &  1.88 &   $+$2.62 &  0.3426 \\
 48055.7545\dotfill &   1990.4470 &  \phn$-$4.28 &  2.09 &   $+$2.95 &  0.3428 \\
 48091.5726\dotfill &   1990.5450 &  $+$16.73 &  2.30 &   $+$4.64 &  0.4063 \\
 48090.8392\dotfill &   1990.5430 &  $+$14.25 &  2.25 &   $+$2.17 &  0.4050 \\
 48117.3576\dotfill &   1990.6156 &  \phn$+$4.41 &  1.69 &   $-$1.38 &  0.4519 \\
 48117.4115\dotfill &   1990.6158 &  \phn$+$6.76 &  1.64 &   $+$0.97 &  0.4520 \\
 48146.2674\dotfill &   1990.6948 &  \phn$+$5.40 &  1.56 &   $-$1.98 &  0.5031 \\
 48146.2833\dotfill &   1990.6948 &  \phn$+$7.56 &  1.79 &   $+$0.18 &  0.5031 \\
 48181.8115\dotfill &   1990.7921 &  $+$11.30 &  1.65 &   $-$0.27 &  0.5660 \\
 48181.8115\dotfill &   1990.7921 &  \phn$+$9.34 &  2.13 &   $-$2.23 &  0.5660 \\
 48204.8156\dotfill &   1990.8551 &  \phn$-$6.40 &  1.67 &   $+$0.62 &  0.6068 \\
 48204.8768\dotfill &   1990.8552 &  \phn$-$7.79 &  1.86 &   $-$0.77 &  0.6069 \\
 48240.9020\dotfill &   1990.9539 &  \phn$-$2.13 &  1.63 &   $-$0.35 &  0.6706 \\
 48240.8130\dotfill &   1990.9536 &  \phn$-$3.20 &  1.99 &   $-$1.44 &  0.6705 \\
 48267.0324\dotfill &   1991.0254 &  $-$10.06 &  1.59 &   $-$0.90 &  0.7169 \\
 48266.8380\dotfill &   1991.0249 &  $-$10.97 &  1.54 &   $-$1.80 &  0.7166 \\
 48293.4845\dotfill &   1991.0978 &  $-$10.81 &  1.88 &   $-$1.03 &  0.7637 \\
 48293.6012\dotfill &   1991.0982 &  $-$12.08 &  2.08 &   $-$2.30 &  0.7639 \\
 48326.1520\dotfill &   1991.1873 &  \phn$+$2.59 &  1.58 &   $-$0.63 &  0.8216 \\
 48326.0668\dotfill &   1991.1870 &  \phn$+$3.56 &  1.88 &   $+$0.34 &  0.8214 \\
 48344.1840\dotfill &   1991.2366 &  \phn$-$5.55 &  1.72 &   $-$1.48 &  0.8535 \\
 48344.1670\dotfill &   1991.2366 &  \phn$-$4.04 &  2.17 &   $+$0.03 &  0.8535 \\
 48382.0545\dotfill &   1991.3403 &  \phn$+$4.77 &  2.41 &   $-$0.87 &  0.9205 \\
 48381.9791\dotfill &   1991.3401 &  \phn$+$3.36 &  3.12 &   $-$2.28 &  0.9204 \\
 48397.5598\dotfill &   1991.3828 &  \phn$+$4.72 &  1.70 &   $-$0.71 &  0.9480 \\
 48397.5052\dotfill &   1991.3826 &  \phn$+$4.29 &  1.81 &   $-$1.14 &  0.9479 \\
 48435.2521\dotfill &   1991.4860 &  \phn$-$2.14 &  1.91 &   $+$2.33 &  0.0147 \\
 48435.2451\dotfill &   1991.4860 &  \phn$-$2.18 &  2.14 &   $+$2.29 &  0.0147 \\
 48457.4053\dotfill &   1991.5466 &  \phn$-$2.00 &  2.09 &   $+$2.51 &  0.0539 \\
 48457.4313\dotfill &   1991.5467 &  \phn$-$3.33 &  3.00 &   $+$1.18 &  0.0540 \\
 48489.2887\dotfill &   1991.6339 &  \phn$+$4.97 &  1.84 &   $+$0.88 &  0.1104 \\
 48489.3417\dotfill &   1991.6341 &  \phn$+$3.04 &  1.50 &   $-$1.05 &  0.1105 \\
 48521.2908\dotfill &   1991.7215 &  \phn$+$0.32 &  1.60 &   $-$1.21 &  0.1670 \\
 48521.3087\dotfill &   1991.7216 &  \phn$+$2.00 &  1.90 &   $+$0.47 &  0.1671 \\
 48583.7983\dotfill &   1991.8927 &  \phn$+$6.38 &  2.10 &   $-$3.46 &  0.2777 \\
 48583.9000\dotfill &   1991.8930 &  \phn$+$9.65 &  2.18 &   $-$0.20 &  0.2779 \\
 48607.5708\dotfill &   1991.9578 &  \phn$-$6.31 &  2.11 &   $+$0.34 &  0.3198 \\
 48607.3726\dotfill &   1991.9572 &  \phn$-$5.34 &  2.21 &   $+$1.30 &  0.3195 \\
 48607.7433\dotfill &   1991.9582 &  \phn$-$7.45 &  2.17 &   $-$0.61 &  0.3201 \\
 48607.7876\dotfill &   1991.9584 &  $-$10.48 &  2.22 &   $-$3.64 &  0.3202 \\
 48638.0588\dotfill &   1992.0412 &  \phn$-$0.21 &  1.78 &   $-$1.53 &  0.3738 \\
 48638.0004\dotfill &   1992.0411 &  \phn$-$0.82 &  2.02 &   $-$2.15 &  0.3737 \\
 48668.1421\dotfill &   1992.1236 &  $-$15.15 &  1.68 &   $-$3.91 &  0.4270 \\
 48668.1951\dotfill &   1992.1237 &  $-$13.20 &  1.75 &   $-$1.96 &  0.4271 \\
 48725.4635\dotfill &   1992.2805 &  \phn$-$2.79 &  1.84 &   $-$2.39 &  0.5285 \\
 48725.3516\dotfill &   1992.2802 &  \phn$-$2.23 &  2.39 &   $-$1.84 &  0.5283 \\
 48740.8474\dotfill &   1992.3226 &  $-$10.82 &  1.60 &   $-$0.68 &  0.5558 \\
 48740.8534\dotfill &   1992.3227 &  $-$10.36 &  1.55 &   $-$0.21 &  0.5558 \\
 48798.0636\dotfill &   1992.4793 &  \phn$+$1.41 &  1.80 &   $-$1.43 &  0.6571 \\
 48798.1008\dotfill &   1992.4794 &  \phn$+$1.61 &  1.87 &   $-$1.23 &  0.6571 \\
 48832.6704\dotfill &   1992.5740 &  \phn$+$3.87 &  3.15 &   $-$3.40 &  0.7183 \\
 48832.7720\dotfill &   1992.5743 &  \phn$+$2.58 &  2.84 &   $-$4.68 &  0.7185 \\
 48947.7079\dotfill &   1992.8890 &  \phn$-$5.01 &  1.95 &   $-$0.73 &  0.9220 \\
 48947.6419\dotfill &   1992.8888 &  \phn$-$3.46 &  3.21 &   $+$0.83 &  0.9219 \\
 48947.9211\dotfill &   1992.8896 &  \phn$-$3.18 &  2.38 &   $+$0.97 &  0.9224 \\
 48947.9711\dotfill &   1992.8897 &  \phn$+$1.57 &  2.77 &   $+$5.72 &  0.9225 \\
 49023.4580\dotfill &   1993.0964 &  \phn$+$9.02 &  2.36 &   $+$3.96 &  0.0561 \\
 49023.4832\dotfill &   1993.0965 &  \phn$+$5.99 &  4.55 &   $+$0.93 &  0.0561 \\
 49056.5632\dotfill &   1993.1870 &  \phn$-$4.23 &  1.34 &   $+$1.55 &  0.1147 \\
 49056.5602\dotfill &   1993.1870 &  \phn$-$4.68 &  1.11 &   $+$1.10 &  0.1147 \\
\enddata
\tablenotetext{a}{Abscissa residuals as provided in the original {\it Hipparcos\/} 5-parameter solution (see text).}
\tablenotetext{b}{Original uncertainties have been scaled by the factor 0.84 (see text).}
\end{deluxetable}


\vskip 1in

\begin{deluxetable}{lccccc}
\tabletypesize{\scriptsize}
\tablewidth{0pc}
\tablecaption{Radial velocity measurements of \phiher\ from
\cite{Aikman:76}.\label{tab:aikman}}
\tablehead{\colhead{HJD} & \colhead{} & \colhead{RV} & \colhead{$\sigma_{\rm RV}$\tablenotemark{a}} & \colhead{$O\!-\!C$} & \colhead{} \\
\colhead{\hbox{~~(2,400,000$+$)~~}} & \colhead{Julian Year} & \colhead{(\kms)} & \colhead{(\kms)} & \colhead{(\kms)} & \colhead{Phase}}
\startdata
 38862.9364\dotfill &  1965.2784 &  $-$13.66 &    1.06 &  $+$0.39 &   0.0676 \\
 38910.7884\dotfill &  1965.4094 &  $-$16.10 &    0.60 &  $-$0.04 &   0.1523 \\
 39328.7297\dotfill &  1966.5537 &  $-$15.94 &    0.21 &  $+$0.14 &   0.8922 \\
 39637.8759\dotfill &  1967.4001 &  $-$17.96 &    0.37 &  $-$0.11 &   0.4396 \\
 39657.7890\dotfill &  1967.4546 &  $-$17.64 &    0.34 &  $+$0.27 &   0.4748 \\
 39658.8948\dotfill &  1967.4576 &  $-$17.63 &    0.32 &  $+$0.28 &   0.4768 \\
 39897.0577\dotfill &  1968.1097 &  $-$16.03 &    0.14 &  $-$0.10 &   0.8984 \\
 39962.9980\tablenotemark{b}\dotfill &  1968.2902 &  $-$13.85 &    0.21 &  ($-$1.37) &   0.0152 \\
 39986.7324\dotfill &  1968.3552 &  $-$13.59 &    0.27 &  $+$0.10 &   0.0572 \\
 39987.7414\dotfill &  1968.3580 &  $-$13.87 &    0.14 &  $-$0.11 &   0.0590 \\
 39988.7903\dotfill &  1968.3608 &  $-$14.05 &    0.18 &  $-$0.23 &   0.0608 \\
 39989.7716\dotfill &  1968.3635 &  $-$13.74 &    0.19 &  $+$0.14 &   0.0626 \\
 40006.8033\dotfill &  1968.4101 &  $-$14.77 &    0.23 &  $+$0.05 &   0.0927 \\
 40041.9075\dotfill &  1968.5062 &  $-$16.23 &    0.21 &  $-$0.13 &   0.1549 \\
 40066.7294\dotfill &  1968.5742 &  $-$16.27 &    0.30 &  $+$0.38 &   0.1988 \\
 40288.0550\dotfill &  1969.1802 &  $-$18.04 &    0.67 &  $-$0.07 &   0.5907 \\
 40356.8535\dotfill &  1969.3685 &  $-$17.70 &    0.14 &  $+$0.08 &   0.7125 \\
 40357.9518\dotfill &  1969.3715 &  $-$17.50 &    0.21 &  $+$0.28 &   0.7144 \\
 40403.8423\dotfill &  1969.4972 &  $-$17.43 &    0.16 &  $-$0.06 &   0.7956 \\
 40407.8790\dotfill &  1969.5082 &  $-$17.18 &    0.53 &  $+$0.13 &   0.8028 \\
 40745.8653\dotfill &  1970.4336 &  $-$18.29 &    0.51 &  $-$0.53 &   0.4012 \\
 40763.8147\dotfill &  1970.4827 &  $-$18.59 &    0.50 &  $-$0.75 &   0.4330 \\
 41026.9650\dotfill &  1971.2032 &  $-$16.03 &    0.51 &  $-$0.11 &   0.8988 \\
 41134.7499\dotfill &  1971.4983 &  $-$15.36 &    0.39 &  $-$0.63 &   0.0897 \\
 41149.7702\dotfill &  1971.5394 &  $-$14.93 &    0.18 &  $+$0.47 &   0.1163 \\
 41214.7197\dotfill &  1971.7172 &  $-$17.01 &    0.48 &  $-$0.06 &   0.2313 \\
 41522.8695\dotfill &  1972.5609 &  $-$17.45 &    0.25 &  $+$0.04 &   0.7768 \\
 41527.7210\dotfill &  1972.5742 &  $-$17.58 &    0.19 &  $-$0.14 &   0.7854 \\
 41554.6734\dotfill &  1972.6480 &  $-$16.93 &    0.19 &  $+$0.10 &   0.8331 \\
 41632.5999\dotfill &  1972.8613 &  $-$13.11 &    0.25 &  $+$0.19 &   0.9711 \\
 41735.0464\dotfill &  1973.1418 &  $-$16.96 &    0.34 &  $-$0.90 &   0.1525 \\
 41750.0084\dotfill &  1973.1828 &  $-$16.88 &    0.42 &  $-$0.45 &   0.1789 \\
 41775.9408\dotfill &  1973.2538 &  $-$17.82 &    0.62 &  $-$0.92 &   0.2249 \\
 41793.8875\dotfill &  1973.3029 &  $-$17.23 &    0.65 &  $-$0.09 &   0.2566 \\
 41883.7190\dotfill &  1973.5489 &  $-$18.00 &    0.58 &  $-$0.20 &   0.4157 \\
 41898.7083\dotfill &  1973.5899 &  $-$17.78 &    0.21 &  $+$0.08 &   0.4422 \\
 41929.6650\dotfill &  1973.6746 &  $-$18.00 &    0.25 &  $-$0.06 &   0.4970 \\
\enddata
\tablenotetext{a}{Original uncertainties have been scaled by the factor 1.77 to yield a reduced $\chi^2$ near unity in the combined fit.}
\tablenotetext{b}{Observation excluded from the fit (see text).}
\end{deluxetable}

\clearpage

\begin{deluxetable}{lccccc}
\tabletypesize{\scriptsize}
\tablewidth{0pc}
\tablecaption{Radial velocity measurements of \phiher\ from
\cite{Adelman:01}.\label{tab:adelman}}
\tablehead{\colhead{HJD} & \colhead{} & \colhead{RV} & \colhead{$\sigma_{\rm RV}$\tablenotemark{a}} & \colhead{$O\!-\!C$} & \colhead{} \\
\colhead{\hbox{~~(2,400,000$+$)~~}} & \colhead{Julian Year} & \colhead{(\kms)} & \colhead{(\kms)} & \colhead{(\kms)} & \colhead{Phase}}
\startdata
 47751.7740\dotfill &  1989.6147 &  $-$17.00 &    0.47 &  $-$0.12 &   0.8047 \\
 48141.7160\dotfill &  1990.6823 &  $-$17.10 &    0.47 &  $+$0.42 &   0.4950 \\
 48705.5950\tablenotemark{b}\dotfill &  1992.2261 &  $-$16.00 &    0.47 &  ($+$1.52) &   0.4933 \\
 48706.9190\dotfill &  1992.2298 &  $-$17.20 &    0.47 &  $+$0.32 &   0.4957 \\
 49134.9660\tablenotemark{b}\dotfill &  1993.4017 &  $-$22.00 &    0.47 &  ($-$5.30) &   0.2535 \\
 49394.9930\dotfill &  1994.1136 &  $-$17.80 &    0.47 &  $-$0.44 &   0.7139 \\
 50166.9250\dotfill &  1996.2270 &  $-$14.20 &    0.47 &  $-$0.15 &   0.0805 \\
 50168.0620\dotfill &  1996.2301 &  $-$13.30 &    0.47 &  $+$0.81 &   0.0825 \\
 50169.0250\dotfill &  1996.2328 &  $-$13.50 &    0.47 &  $+$0.66 &   0.0843 \\
 50591.7310\dotfill &  1997.3901 &  $-$16.80 &    0.47 &  $-$0.18 &   0.8326 \\
 50592.9870\dotfill &  1997.3935 &  $-$16.50 &    0.47 &  $+$0.09 &   0.8348 \\
 50593.9840\dotfill &  1997.3963 &  $-$16.80 &    0.47 &  $-$0.23 &   0.8366 \\
 50595.9090\dotfill &  1997.4015 &  $-$16.70 &    0.47 &  $-$0.17 &   0.8400 \\
 50653.9980\dotfill &  1997.5606 &  $-$14.80 &    0.47 &  $-$0.71 &   0.9429 \\
 50654.9910\dotfill &  1997.5633 &  $-$14.70 &    0.47 &  $-$0.68 &   0.9446 \\
 50655.8160\dotfill &  1997.5655 &  $-$14.00 &    0.47 &  $-$0.04 &   0.9461 \\
 50657.0010\dotfill &  1997.5688 &  $-$13.90 &    0.47 &  $-$0.03 &   0.9482 \\
 50658.9960\dotfill &  1997.5743 &  $-$12.90 &    0.47 &  $+$0.82 &   0.9517 \\
 50943.8490\dotfill &  1998.3541 &  $-$17.30 &    0.47 &  $+$0.17 &   0.4560 \\
 51028.8540\dotfill &  1998.5869 &  $-$18.10 &    0.47 &  $-$0.55 &   0.6065 \\
\enddata
\tablenotetext{a}{Uncertainties were not reported in the original publication, so the values listed here were derived from the requirement that the reduced $\chi^2$ for the velocities be near unity in the combined fit.}
\tablenotetext{b}{Observation excluded from the fit (see text).}
\end{deluxetable}

\vskip 1in

\begin{deluxetable}{lcccc}
\tablewidth{0pc}
\tablecaption{Observed and predicted properties of \phiher.\label{tab:colors}}
\tablehead{ \colhead{} & \colhead{Mass} & \colhead{$M_V$} & \colhead{$U\!-\!B$} & \colhead{$B\!-\!V$} \\
\colhead{~~~~~~~~Object~~~~~~~~} & \colhead{(M$_{\sun}$)} & \colhead{(mag)} & \colhead{(mag)} & \colhead{(mag)}}
\startdata
\noalign{\vskip 2pt}
\multicolumn{5}{c}{Observed} \\
\noalign{\vskip 3pt}
\hline
\noalign{\vskip 3pt}
Primary\dotfill    & 3.05~$\pm$~0.24   & 0.100~$\pm$~0.059 & \nodata & \nodata \\
Secondary\dotfill  & 1.614~$\pm$~0.066 & 2.670~$\pm$~0.074 & \nodata & \nodata \\
Combined\dotfill   & \nodata           & \nodata           & $-$0.250~$\pm$~0.009\phs & $-$0.068~$\pm$~0.008\phs \\
\noalign{\vskip 3pt}
\hline
\noalign{\vskip 4pt}
\multicolumn{5}{c}{Predicted for [Fe/H] $= -0.03$ and Age = 210 Myr} \\
\noalign{\vskip 3pt}
\hline
\noalign{\vskip 3pt}
Primary\dotfill    & 3.09    & 0.098    & $-$0.267   & $-$0.094 \\
Secondary\dotfill  & 1.590   & 2.669    & +0.053     & +0.271   \\
Combined\dotfill   & \nodata & \nodata  & $-$0.249   & $-$0.067 \\
\enddata
\tablecomments{Predictions are based on Yonsei-Yale models by
\cite{Yi:01} \citep[see also][]{Demarque:04}.}
\end{deluxetable}

\clearpage

\clearpage

\clearpage

\clearpage

\clearpage

\clearpage

\clearpage

\end{document}